\documentclass[aps,prl,twocolumn,showpacs,superscriptaddress]{revtex4}
\usepackage[english]{babel}
\usepackage{graphicx}
\usepackage{amsmath}
\usepackage{amssymb}
\usepackage{natbib}
\usepackage{times}
\usepackage{soul}

\begin{document}

\title{Information transfer by vector spin chirality in finite
magnetic chains}

\author{Matthias Menzel}
\affiliation{Institut f\"ur Angewandte Physik, Universit\"at Hamburg,
Jungiusstr.~11, 20355 Hamburg, Germany}

\author{Yuriy Mokrousov}
\affiliation{Peter Gr\"unberg Institut and Institute for
Advanced Simulation, Forschungszentrum J\"{u}lich, D-52425
J\"{u}lich, Germany}

\author{Robert Wieser}
\affiliation{Institut f\"ur Angewandte Physik, Universit\"at Hamburg,
Jungiusstr.~11, 20355 Hamburg, Germany}

\author{Jessica E. Bickel}
\affiliation{Institut f\"ur Angewandte Physik, Universit\"at Hamburg,
Jungiusstr.~11, 20355 Hamburg, Germany}

\author{Elena Vedmedenko}
\affiliation{Institut f\"ur Angewandte Physik, Universit\"at Hamburg,
Jungiusstr.~11, 20355 Hamburg, Germany}

\author{Stefan Bl\"ugel}
\affiliation{Peter Gr\"unberg Institut and Institute for
Advanced Simulation, Forschungszentrum J\"{u}lich, D-52425
J\"{u}lich, Germany}

\author{Stefan Heinze}
\affiliation{Institut f\"ur Theoretische Physik und Astrophysik,
Christian-Albrecht-Universit\"at zu Kiel, Leibnizstr. 15, 24098 Kiel, Germany}

\author{Kirsten von Bergmann}
\affiliation{Institut f\"ur Angewandte Physik, Universit\"at Hamburg,
Jungiusstr.~11, 20355 Hamburg, Germany}

\author{Andr\'e Kubetzka}
\affiliation{Institut f\"ur Angewandte Physik, Universit\"at Hamburg,
Jungiusstr.~11, 20355 Hamburg, Germany}

\author{Roland Wiesendanger}
\affiliation{Institut f\"ur Angewandte Physik, Universit\"at Hamburg,
Jungiusstr.~11, 20355 Hamburg, Germany}

\date{\today}

\begin{abstract}
Vector spin chirality is one of the fundamental characteristics of complex 
magnets. For a one-dimensional spin-spiral state it can be interpreted 
as the handedness, or rotational sense of the spiral. Here, using 
spin-polarized scanning tunneling microscopy, we demonstrate the occurrence 
of an atomic-scale spin-spiral in finite individual bi-atomic Fe chains on 
the (5$\times$1)-Ir(001) surface. We show that the broken inversion symmetry 
at the surface promotes one direction of the vector spin chirality, leading 
to a unique rotational sense of the spiral in all chains. Correspondingly, 
changes in the spin direction of one chain end can be probed tens of nanometers 
away, suggesting a new way of transmitting information about the state of 
magnetic objects on the nanoscale.
\end{abstract}


\maketitle

The concept of the vector spin chirality (VSC) in an ensemble of spins 
proved to be remarkably fruitful in the fields of magnetism of frustrated 
systems~\cite{Villain1977,Grohol2005, Zee1989} and 
multiferroics~\cite{Mostovoy2007}, where it is naturally coupled to the 
electric polarization~\cite{Balatsky2005}. For a one-dimensional spin 
spiral the VSC is given by 
$\boldsymbol{\kappa}_{i,i+1}\propto\mathbf{S}_i\times\mathbf{S}_{i+1}$ 
with the spins $\mathbf{S}_i$ and $\mathbf{S}_{i+1}$ at neighboring sites 
$i$ and $i+1$, and it can be directly related to the spin current flowing 
between this pair of non-collinear spins~\cite{Balatsky2005}. Such a spin 
current transmits information about the direction of $\mathbf{S}_n$ to any 
other atom $m$ of the chain and thereby defines the direction of 
$\mathbf{S}_m$, which can then be read out. A finite one-dimensional spin 
spiral can therefore be used to efficiently probe and transmit 
information about the magnetic state of even an atomic-scale object when it 
interacts with such an object at one end. Practical realization of this 
scenario for information transport would be of great benefit, since it employs 
only the spin degree of freedom and no flow of charge is necessary. 
There are important advantages of using a system with a non-zero VSC over a 
chain with collinear magnetic order~\cite{Hirjibehedin2006,Wiebe2011}: 
First, the VSC is a more robust quantity than the local magnetization with 
respect to temperature and quantum fluctuations typical for one-dimensional 
systems~\cite{Cinti2008}. Further, a spin-spiral state is expected to be 
much less sensitive to parasitic magnetic stray fields always present in a 
device, and it would not be accompanied by creation of domain walls, harmful 
for efficient restructuring of the spin state. From the point of view of 
technological applications, the proposed way of information processing would 
open new paradigms in spintronic-based devices.

\begin{figure}[t]
	\includegraphics[width=7.6cm]{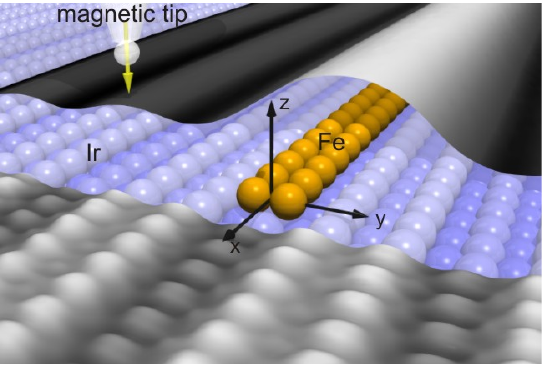}
	\caption{\label{Menzel_fig1} Schematic representation of the investigated 
	sample system and the experimental setup. The STM image in the front shows 
	the atomically resolved (5$\times$1)-reconstructed Ir(001) surface 
	(measurement parameters: $U = -2$\,mV, $I = 40$\,nA, room temperature). 
	Fe atoms deposited at room temperature form bi-atomic chains in the 
	trenches of the reconstruction (ball model), which appear as a single 
	strand in an STM image as shown in the overlaid image at the top 
	(measurement parameters: $U = +400$\,mV, $I = 5$\,nA, $T = 8$\,K). The 
	experimental setup allows to apply magnetic fields perpendicular to the 
	sample surface,~i.e.~along the $z$-axis. To investigate the magnetic 
	order of the Fe chains we use spin-polarized STM with magnetic tips.}
\end{figure}

\begin{figure}[ht]
	\includegraphics[width=8.6cm]{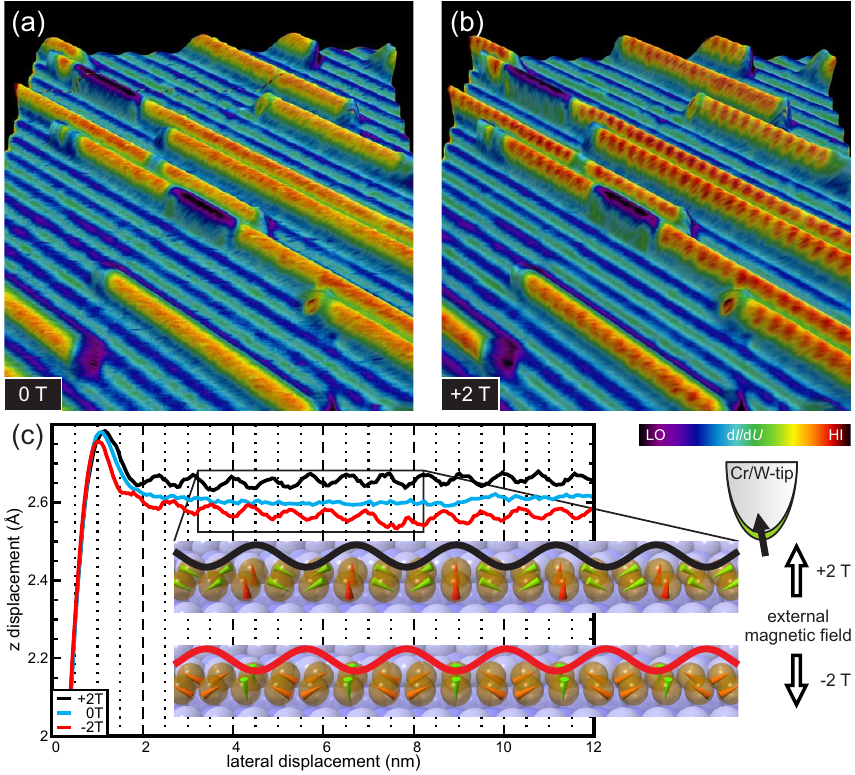}
	\caption{\label{Menzel_fig2} SP-STM measurements of Fe chains on Ir(001). 
	(a) and (b) Typical sample area of $30\times30$\,nm$^2$ measured with an 
	Fe-coated W tip without and with an applied external magnetic 
	field of $B=+2$\,T perpendicular to the sample surface, respectively 
	(constant current images colorized with simultaneously acquired 
	d{\em I}/d{\em U} maps, measurement parameters: 
	$U=+500$\,mV, $I=5$\,nA, $T=8$\,K). (c) Topographic line profiles of the 
	same Fe chain at $B=0$\,T and $B=\pm2$\,T measured with a Cr-coated tip. 
	The insets show schematically the tip magnetization and how a 120$^{\circ}$ 
	spin-spiral, which is inverting in opposite fields, could explain the 
	experimental results.
	}
\end{figure}

In order to realize this visionary goal, we have chosen an appropriate 
model-type system, namely bi-atomic Fe chains with a length of up to 30\,nm, 
which form by self-assembly on a (5$\times$1)-reconstructed Ir(001) 
surface~\cite{Hammer2003}. They appear as single strands in scanning tunneling 
microscopy (STM) images, reflecting the very small distance 
between the pairs of Fe atoms perpendicular to the chain axis (see 
Fig.~\ref{Menzel_fig1})~\cite{EPAPS}. As shown in Fig.~\ref{Menzel_fig2}(a) 
in a spin-polarized (SP-) STM image~\cite{Wiesendanger2009} in zero external 
magnetic field, the chains appear featureless. The same observation is made for 
spin-averaged STM measurements using non-magnetic W tips not only at $B=0$\,T 
but also in an external magnetic field~\cite{EPAPS}. This uniform appearance of 
the chains changes significantly when examined with a magnetic tip in an 
external magnetic field: all the chains, regardless of their length, exhibit a 
modulation along their axes with a periodicity of three atomic distances 
(see Fig.~\ref{Menzel_fig2}(b)). This image is obtained using a tip sensitive 
to the out-of-plane component of the magnetization. Maxima and minima along the 
chain axis therefore represent areas with magnetization components pointing up 
or down with respect to the surface~\cite{Wortmann2001}. The modulation 
persists over a wide range of applied bias voltages without changing its 
periodicity and could be explained by a spin-spiral state with an angle of 
120$^\circ$ between the magnetic moments of neighboring atoms along the chain 
axis, as it is sketched in Fig.~\ref{Menzel_fig2}(c). The periodic contrast is 
reversed by inverting the external field direction (Fig.~\ref{Menzel_fig2}(c)) 
when a Cr-coated tip is used for imaging. Due to the antiferromagnetic 
ordering of the Cr-coated tip, its magnetization direction does not change 
when applying magnetic fields of this strength~\cite{Pietzsch2004}. 
Thus, the inversion of magnetic contrast along the chain axis can only be 
explained when the spin-spiral structure of the Fe chains exhibits a net 
magnetic moment that aligns with the magnetic field. When turning off the 
magnetic field the modulation vanishes and the chains again appear featureless 
(blue line in Fig.~\ref{Menzel_fig2}(c)). 

\begin{figure}[t!]
	\includegraphics[width=8.6cm]{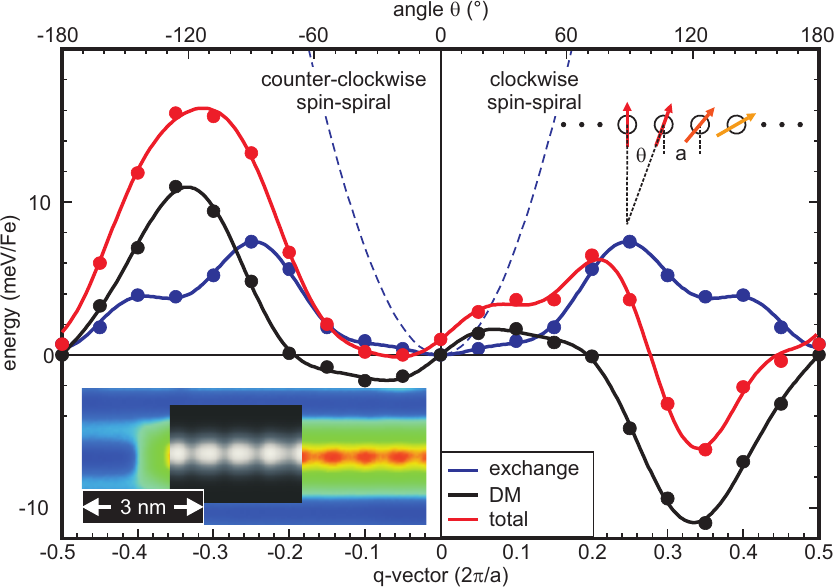}
	\caption{\label{Menzel_fig3} DFT calculations and comparison to 
	experimental findings. Spin-spiral dispersion from first-principles 
	calculations: contributions from Heisenberg exchange (blue dots) and 
	Dzyaloshinskii-Moriya 
	interaction (black dots) and their sum (red dots). The lines represent 
	fits to the calculations with an extended Heisenberg model and the dashed 
	blue line is the Heisenberg exchange dispersion of a free-standing 
	bi-atomic Fe chain. Inset: comparison of a simulated SP-STM image of a 
	120$^\circ$ spin-spiral (gray scale inset; 5\,\AA~above the surface, 
	50\% spin-polarization of the tip) with experimental data (color; constant 
	current image, measurement parameters: $U=+300$\,mV, $I=5$\,nA, $T=8$\,K).}
\end{figure}

In order to verify whether the proposed spin-spiral is the magnetic ground 
state of bi-atomic Fe chains on Ir(001), we performed density functional 
theory (DFT) calculations~\cite{EPAPS}. We scan the magnetic phase space by 
calculating flat spin-spirals which are the general solution of the classical 
Heisenberg model 
$E_{\mathrm H}=-\sum_{ij}J_{ij}{\mathbf S}_i\cdot{\mathbf S}_j$ for a periodic 
lattice with the exchange constant $J_{ij}$ between the spins ${\mathbf S}_i$
and ${\mathbf S}_j$ at the atomic sites $i$ and $j$. Such a spin-spiral, 
propagating along the chain, is given by 
${\mathbf S}_i=S(\cos(qai), 0, \sin(qai))$ where $a$ is the 
lattice constant and ${\mathbf q}=(q,0,0)$ is the characteristic spin-spiral 
vector. Varying $q$ from $q=0$ (ferromagnetic (FM) state) to 
$q=\pm 0.5\,\frac{2\pi}{a}$ (antiferromagnetic (AFM) state), we cover all 
possible spin-spirals and calculate the spin-spiral dispersion energy $E(q)$ 
(blue dots in Fig.~\ref{Menzel_fig3}). Positive and negative values of $q$ 
denote clockwise and counter-clockwise spin-spirals, respectively, and the 
dispersion is symmetric with respect to $\pm q$, as expected from the Coulomb 
interaction. We find that the FM solution is most favorable among 
all states. However, the strong FM exchange found in free-standing bi-atomic Fe
chains (see blue dashed line, FM state $\approx 75$\,meV/Fe~atom below the AFM 
state~\cite{Mokrousov2009}) is almost completely quenched. This is due to the 
strong hybridization with the Ir substrate resulting in a difference between 
the FM and the AFM state of only $\approx1$\,meV/Fe~atom.

Since the exchange interactions in these bi-atomic Fe chains are very small, 
we anticipate a strong influence of the spin-orbit interaction (SOI) on the 
spin-spiral dispersion as proposed in Ref.~\cite{Mazzarello2009}. The SOI 
gives rise to the magnetic anisotropy energy (MAE) 
$E_\mathrm{ani}=\sum_{i}K_i\sin^2\varphi_i$ with the anisotropy constant 
$K_i$ and the angle $\varphi_i$ between the spin ${\mathbf S}_i$ and the easy 
axis at site $i$. The easy axis of the Fe spin moments is out-of-plane 
($z$-axis in Fig.~\ref{Menzel_fig1}) and about 2\,meV/Fe~atom lower in energy 
than the two high-symmetry in-plane directions. Additionally, the SOI induces 
the anti-symmetric Dzyaloshinskii-Moriya (DM) interaction with 
$E_\mathrm{DM}=\sum_{i,j}{\mathbf D}_{ij}\cdot({\mathbf S}_i\times{\mathbf S}_j)$,
where $\mathbf{D}_{ij}$ is the DM vector. At surfaces a non-vanishing 
$E_\mathrm{DM}$ is always possible due to inversion-asymmetry~\cite{Bode2007}. 
If the DM interaction can compete with the Heisenberg exchange it can give rise 
to a unique direction of the VSC. Based on symmetry arguments, 
$\mathbf{D}_{ij}$ and thus the VSC aligns along the $y$-axis, favoring 
cycloidal spin-spirals with a unique rotational sense where the magnetic 
moments rotate in the $xz$-plane (cf. 
Fig.~\ref{Menzel_fig1})~\cite{Dzyaloshinskii1958,Moriya1960,Heide2006}.

We evaluate the correction to $E(q)$ due to the DM interaction, 
$E_{\rm DM}(q)$, from DFT~\cite{EPAPS}. The results for $E_{\rm DM}(q)$ are 
presented as black dots in Fig.~\ref{Menzel_fig3}, and show that the magnitude 
of $E_{\rm DM}(q)$ competes with the contribution from the Heisenberg exchange, 
reaching as much as $10$\,meV. Summing up Heisenberg and DM contributions and 
including an energy shift due to the MAE, we find a robust cycloidal 
spin-spiral ground state, several meV below the FM state, in the vicinity of 
$q\approx+1/3\,\frac{2\pi}{a}$ (red dots). This value and sign of the 
spin-spiral vector corresponds to a clockwise 120$^{\circ}$ spin-spiral, 
running along the chain axis. Owing to the anti-symmetric nature of 
$E_{\rm DM}(q)$ with respect to $\pm q$, the counter-clockwise 
spin-spiral state with $q=-1/3\,\frac{2\pi}{a}$ is much higher in energy, 
which manifests the unique rotational sense of our spin 
structure~\cite{Heide2006} described by a unique VSC that points along the 
$y$-axis for all chains. Remarkably, the magnetic unit cell of this 
120$^{\circ}$-state consists of three Fe atoms along the chain axis, which is 
the minimal periodicity necessary to form a non-collinear state. As shown in 
Fig.~\ref{Menzel_fig3}, a simulated SP-STM image of such a state based on the 
DFT calculations is in excellent agreement with the experimental findings.

The first-principles calculations correspond to a situation of Fe chains 
at zero temperature and zero magnetic field, and in order to understand the 
influence of temperature and magnetic field on our system we use a heat-bath 
MC method~\cite{Wieser2008} employing an extended Heisenberg 
model $E_\mathrm{tot}=E_\mathrm{H}+E_\mathrm{ani}+E_\mathrm{DM}+E_\mathrm{B}$
including the effect of an out-of-plane magnetic field 
$E_\mathrm{B}=-\mu_sB\sum_iS_i^z$, where $\mu_s$ is the magnetic moment. The 
material parameters $J_{ij}$ and ${\mathbf D}_{ij}$ are obtained from fits to 
the first-principles calculations 
(see blue and black line in Fig.~\ref{Menzel_fig3}). 
To capture the non-trivial dispersions of $E(q)$ and $E_{\rm DM}(q)$, we 
included the $J_{ij}$ and ${\mathbf D}_{ij}$ 
parameters up to six nearest neighbors along the chain axis, 
which are listed in the Supplemental Material~\cite{EPAPS}. 
The coupling between the pair of Fe atoms along the $y$ axis is FM with 
$J_{\mathrm{perp}}\approx160$\,meV/Fe~atom and the spin 
moment of each Fe atom and the MAE were set to their respective DFT values of 
$\mu_s=2.75\,\mu_B$ and $K_i=2$\,meV, irrespective of the position 
of the atom within the finite chain. 
Since the Fe chains observed in the experiments show a broad length 
distribution ranging from only a few nm, corresponding to $\approx$\,10~pairs 
of atoms, up to several tens of nm, corresponding to up to 150~pairs, we chose 
to investigate a representative chain consisting of 30 pairs of atoms. 
The value of the time-averaged magnetization $M_z$ in our 
MC simulations is normalized, meaning that a value of one corresponds to a 
non-fluctuating magnetic moment, while a smaller value indicates the 
presence of thermal fluctuations. 
The analysis of the time-averaged out-of-plane magnetization 
$M_z$ of the entire chain at zero field shows that it is negligible for all 
temperatures, see Fig.~4(a), explaining the lack of magnetic contrast in our 
experiments at $B=0$\,T. At the measurement temperature of 8\,K a single 
snapshot of the $x$-, $y$- and $z$-component of the magnetization along 
the chain reveals a 120$^{\circ}$ clockwise spin-spiral, except for small 
deviations in the angles, but rapid switching of this spin-spiral results in 
a vanishing time-averaged $M_z$.

\begin{figure}[t!]
	\includegraphics[width=8.6cm]{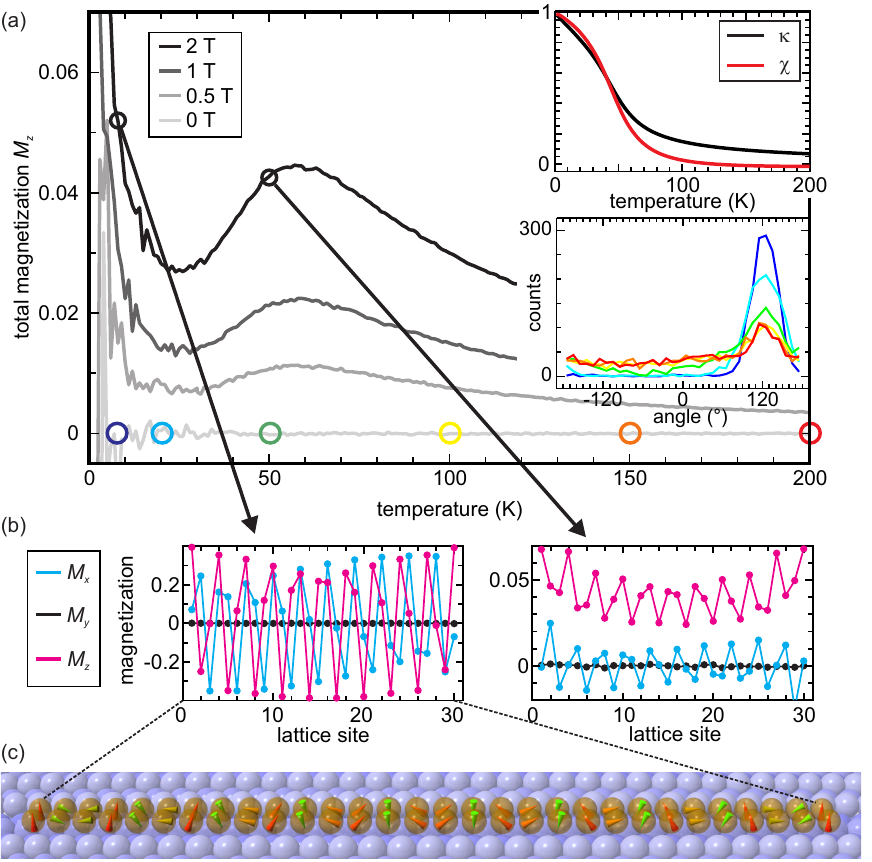}
	\caption{\label{Menzel_fig4} Monte-Carlo simulations for a 30 atom long 
	chain. (a) Temperature dependence of $M_z$ for different external magnetic 
	fields applied along the $z$-axis. Upper inset depicts the temperature 
	dependence of the normalized chiral and scalar order parameters as 
	function of temperature. Lower inset displays the distribution of angles 
	between nearest neighbor spins along the chain (the color of the lines 
	corresponds to the temperature of the system given by the circle of the 
	corresponding color below). (b) Laterally resolved time-averaged 
	magnetization for two different temperatures at $B=+2$\,T. (c) 
	Three-dimensional representation of magnetization components of every 
	atom for $T=8$\,K and $B=+2$\,T.}
\end{figure}

In order to take a deeper look into the intrinsic spin-ordering of the Fe 
chains at $B=0$~T as a function of temperature, we calculate the normalized 
chiral and scalar order parameters, 
$\kappa=\frac{1}{\sin\theta_0}\langle \left(\mathbf{S}_i \times
\mathbf{S}_{i+1} \right)_y \rangle$ and $\chi=\frac{\left\langle
  \mathrm{arccos}\left(\mathbf{S}_i \cdot \mathbf{S}_{i+1}
  \right)\right\rangle - \theta_\infty}{\theta_0 - \theta_\infty}$
 ($\theta_0 = 122.4^\circ$ and $\theta_\infty = 90^\circ$ are the angles of the 
spin configuration at zero and at infinite temperatures), obtained 
from time- and chain-averaged VSC and scalar product between neighboring 
spins, respectively~\cite{Cinti2011}. For low temperatures, see Fig.~4(a) 
upper inset, both $\kappa$ and $\chi$ are close to one, which corresponds to an 
almost ideal clockwise 120$^{\circ}$ spin-spiral state at any time. With 
increasing temperature the local fluctuations of the spin moments of each atom 
become more and more pronounced, which results in a state with $\chi\approx0$ 
at temperatures above 100\,K. Surprisingly, at this temperature the chiral 
order parameter $\kappa$ is still far from zero, and it decays much more 
slowly with temperature. This means that the chiral correlation between the 
spins is still present for $T\gg 100$\,K $-$ a situation reminiscent of that 
for the vector spin chiral liquid state, predicted to occur in one-dimensional 
spin systems~\cite{Hikihara2008,Sudan2009,Furukawa2010,Cinti2011} and observed 
experimentally in the quasi-one-dimensional molecular helimagnetic compound 
Gd($hfac$)$_3$NITEt~\cite{Cinti2008}. It can be understood by looking at the 
time- and chain-averaged distribution of the angle between the nearest 
neighbor spins in the $xz$-plane as a function of temperature, lower inset of 
Fig.~4(a). Despite the fact that the averaged nearest-spin scalar product 
becomes negligible very quickly, the noticeable preference of the spin's 
fluctuations with respect to its neighbor towards positive angles remains for 
a very wide range of temperatures. One can imagine that, under conditions of a 
sufficiently strong VSC, such asymmetry can be used to obtain information about 
the spin's fluctuations on one end of the chain by monitoring the dynamics of 
the spin at the other end, even at comparatively high temperatures.

At finite $B$ an out-of-plane magnetization $M_z$ of the chain arises, see 
Fig.~4(a), and at the experimental conditions of $T=8$\,K and $B=2$\,T we find 
a time-averaged magnetization of the atoms in the chain as displayed in 
Fig.~4(b) (left) and (c), in good agreement with the SP-STM measurements. The 
external magnetic field can induce a net moment of the chain,~e.g.~owing to 
the larger susceptibility of the end spins which have a reduced 
coordination number, thus facilitating a preferred orientation of the 
spin-spiral within the chain. Such a finite size mechanism appears to be 
crucial for the experimental observation of 
the non-collinear magnetic state of our chains, since for an infinite chain 
with the same parameters, the averaged local magnetization as determined 
by the MC simulations is always zero at any temperature, 
even for very large magnetic fields. 
For our 30 atom long chain in Fig.~4(a) at low $T$ the net moment is 
preferentially oriented along the magnetic field, resulting in a comparably 
large value of $M_z$. With increasing $T$ this value decreases due to thermal 
fluctuations. For even larger $T$ the spin-spiral order is weakened and all 
spins have components parallel to the external field, thus contributing to 
$M_z$, see Fig.~4(b) (right). Above the transition temperature of roughly 60\,K 
the magnetic order is destroyed by thermal fluctuations.

\begin{figure}[t!]
	\includegraphics[width=8.6cm]{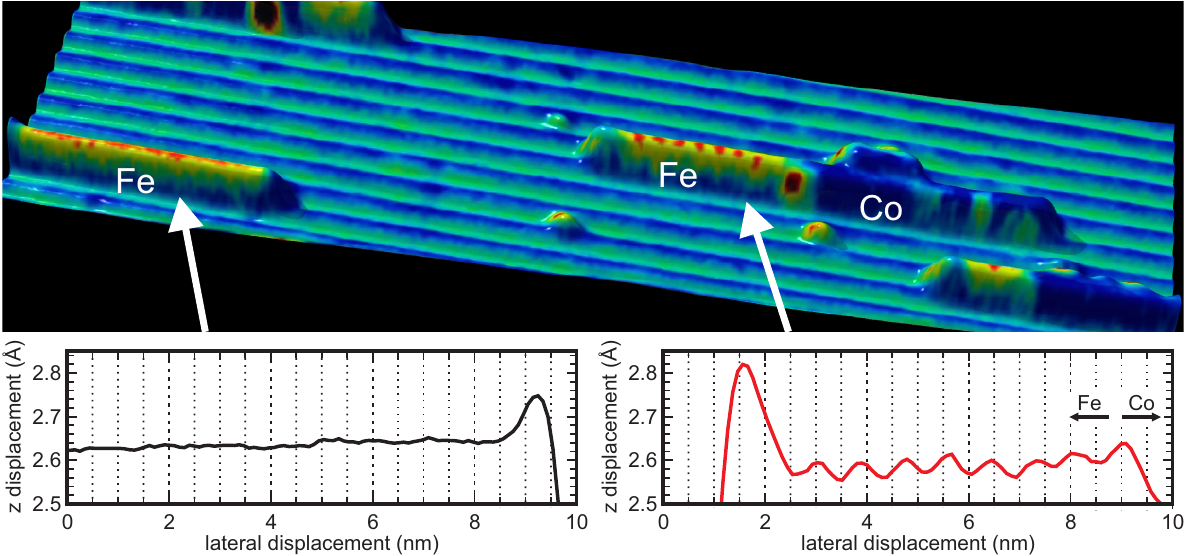}
	\caption{\label{Menzel_fig5} SP-STM measurement of an isolated Fe chain 
	(left) and an Fe chain directly connected to a ferromagnetic Co chain 
	(center) taken at $B=0$\,T with a Cr bulk tip (50$\times$12.5\,nm$^2$ 
	constant current image, colorized with d{\em I}/d{\em U}-map, measurement 
	parameters: $U=+500$\,mV, $I=2$\,nA, $T=8$\,K). As seen in both the image 
	and the line profiles below, the isolated Fe chain appears featureless 
	while the one attached to the Co shows the periodicity indicative of 
	spin-spiral order.}
\end{figure}

Finally, we demonstrate that the suggested novel mechanism of information 
transport based on VSC can be realized in this type of system by attaching 
the Fe chain to a stable magnetic 
particle. Figure~5 shows an SP-STM image at $B = 0$\,T of one isolated 
Fe chain (bottom left) and one Fe chain coupled to a ferromagnetic Co chain 
(center). The isolated chain 
does not show the periodicity indicative of spin-spiral 
order due to rapid thermal switching (cf.~Fig.~2(a)). 
In contrast, the Fe chain that is attached to the Co chain displays a distinct 
magnetic signal. This 
stabilization of the spin-spiral happens due to strong direct exchange 
interaction of the end Fe spins with the Co chain, which fixes their direction 
in space. 
With the magnetization of one chain end fixed in space and time, the 
intra-chain coupling leads to a stabilization of the whole Fe chain.
Correspondingly, by measuring the direction of any Fe spin, even 
nanometers away, we can 'read out' the state of the Co chain and monitor its 
changes, owing to the link by the vector spin chirality. 

{\em Acknowledgements.}
We thank Y.~Yoshida, S.~Schr\"oder, D.~S.~G.~Bauer, Ph.~Mavropoulos,
B.~Zimmermann, M.~Heide, W.~Selke and G.~Bihlmayer for fruitful discussions. 
M.M., R.Wieser, J.E.B., E.V., K.v.B., A.K., R.Wie\-sen\-dan\-ger gratefully 
acknowledge financial support from the DFG via SFB668, from the EU via the ERC 
Advanced Grant FURORE, and from the Cluster of Excellence NANOSPINTRONICS 
funded by the Forschungs- und Wissenschaftsstiftung Hamburg.  J.E.B. also 
acknowledges financial support from the Alexander von Humboldt Foundation. 
Y.M. acknowledges the J\"ulich Supercomputing Centre and funding under the 
HGF-YIG Programme VH-NG-513 and S.H. thanks the DFG for financial support 
under grant number HE3292/8-1.

\end{document}